\documentclass{article}

\usepackage{arxiv}

\usepackage[utf8]{inputenc} 
\usepackage[T1]{fontenc}    
\usepackage{hyperref}       
\usepackage{url}            
\usepackage{booktabs}       
\usepackage{amsfonts}       
\usepackage{nicefrac}       
\usepackage{microtype}      
\usepackage{lipsum}		
\usepackage{graphicx}
\usepackage{doi}
\usepackage[tight]{units}
\def \be {\begin{equation}}
\def \ee {\end{equation}}

\title{Viscoelastic bubbly media and ultrasonic shear-mode effects}

\date{} 					

\author{ \href{https://orcid.org/0000-0003-3442-4445}{\includegraphics[scale=0.06]{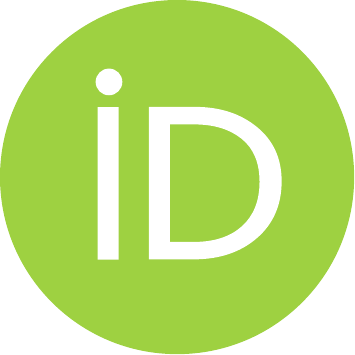}\hspace{1mm}Michael Forrester}\\
	Advanced Materials and Devices\\
	QinetiQ\\
	Farnborough\\ 
	UK\\
	GU14 0LX
	\And
	Valerie Pinfield \\
	Chemical engineering\\
	Loughborough University\\
	LU11 3TU\\
	}

\begin{document}
\maketitle

\begin{abstract}
Here we show that in ultrasonic fields the phenomenon of reconversion of shear-modes into an effective compressional wave has a significant effect for bubbles in a medium viscosity liquid or weak gel. We present the consequent extra terms in the effective wavenumber and find the changes in sound velocity for different bubble radii. At high concentrations of bubbles the inclusion of shear-mode effects in the multiple-scattering model can help identify bubble sizes where additional resonance signatures emerge.
\end{abstract}


Rayleigh gave the first mathematical description of compressible oscillations of bubbles in a liquid in $1917$ \cite{Rayleigh1917} without inclusion of surface tension or viscous effects. The Rayleigh model was later applied to the dynamics of bubble cavitation by Plesset \cite{Plesset1949}. The first consideration of the collapse or growth of a bubble in a viscous fluid was carried out by Poritsky in $1952$ \cite{Poritsky1952}. For low concentrations of bubbles the Foldy model of $1945$ is successful for describing the attenuation and sound velocity \cite{Foldy1945}. However, it does not take into account inter-bubble coupling effects. Importantly, no examination of the high concentration phenomena has been carried out and consequences linked to the recombination of energy fluxes in the compressional wave have been disregarded on dissipation arguments. Historically, the approaches taken to describe bubble acoustics have fallen under two categories: linear and non-linear modelling, where traditionally linear models have formed the initial understanding to develop more complex dissipative arguments. The Foldy model is based upon linear bubble dynamics. In the work of Commander and Prosperetti \cite{Commander1989}, based on the findings by van Wijngaarden \cite{Wijngaarden1968}  and Caflisch et al\cite{Caflisch1985}, linear bubble oscillations were examined against available data. We consider the bubbles to have a monodisperse size distribution and identical initial conditions. Technologies have been developed that can produce bubbles of uniform size distribution and have been reported as having high stability. A stable bubble is more durable against pressure changes and does not implode or expand rapidly. From this perspective a linear model is applicable. Bubbles may dissolve quite rapidly and for this reason small nuclei can be coated in nanoparticles to increase their lifetime\cite{StrideCoussios} and to slow or prevent Ostwald-Ripening and convergence to a larger bubble. Low excitation pressures may lead to a linear approximation to the oscillations.
The bubbles at higher pressures are, however, deformable, sensitive to changes in the continuous phase, and can fuse or fission under perturbation. Bubble growth and collapse can occur in a flow which is important for a range phenomena. The stable bubbles can also be dispersed in a liquid medium. When the number of bubbles becomes larger than a few percent by volume concentration, i.e. there is a high bubble fraction in the continuous phase, some degree of shear-mode conversion back to the primary compressional wave can occur. Multiple scattering formulations have until the present day been based upon the work of Foldy \cite{Foldy1945}, Lax\cite{Lax1951}, Waterman/Truell\cite{WatermanTruell1961}, Fikioris/Waterman\cite{FikiorisWaterman1964}, Epstein/Carhart\cite{EpsteinCarhart1953}, and Allegra/Hawley\cite{AllegraHawley1972}. The latter two form the basis of the $ECAH$ model which was initially derived to investigate the fine particles in fog. A corrected form of the Waterman Truell model was given by Lloyd/Berry ($LB$)\cite{LloydBerry1967} in $1967$ and later re-derived classically by Linton/Martin\cite{LintonMartin2006} in $2006$. A new model was formulated in $2012$ by Lupp\'e/Conoir/Norris ($LCM$) that enables a multiple scattering approach inclusive of thermal and viscous effects\cite{Luppe2012,PinfieldForrester2015}. It is from the $LB$, $ECAH$,  and $LCM$ models that we make an extension to include shear-mode conversion effects.  For contrast agents such as bubbles in a moderately viscous continuous phase these extra terms have significant contributions to the attenuation and phase velocity. The $LCM$ model was applied to compressional waves, extracting first order coefficients, and assuming thermodynamic equilibrium. Shear-mode scattering coefficients were found  by taking the first term in a series expansion in $k_C r$ using the boundary equations of the $ECAH$ model\cite{EpsteinCarhart1953,AllegraHawley1972}, where $k_C$ is the compressional wavenumber and $r$ is the bubble radius. These scattering coefficients are driven by a dependency upon the ratio of the densities of the dispersed bubbles ($\rho$) and a continuous phase ($\rho'$): $\widehat{\rho}=\rho'/\rho$. This can be seen in the following equations where the compressional-compressional ($T_1^{CC}$), compressional-shear ($T_1^{CS}$), shear-compressional ($T_1^{SC}$) and shear-shear ($T_1^{SS}$) coefficients are defined for partial wave order one:

\be
T_1^{CC}=\frac{i \left(k_C r\right)^3 \left(\widehat{\rho}-1\right)h_2 \left(k_S r\right)}{3D\left(k_S r\right)},
\ee
\be
T_1^{CS}=\frac{k_C r\left(\widehat{\rho}-1\right)}{k_S r D\left(k_S r\right)},
\ee
\be
T_1^{SC}=-\frac{2i}{3}\left(k_C r\right)^2 k_S r\frac{\left(\widehat{\rho}-1\right)F\left(k_S r\right)}{D\left(k_S r\right)},
\ee
\be
T_1^{SS}=-\frac{3j_2\left(k_s r\right)-2\left(\widehat{\rho}-1\right) j_0\left(k_s r\right)}{D\left(k_S r\right)},
\ee
with $F\left(k_S r\right)=h_2 \left(k_S r\right)j_0 \left(k_S r\right)-h_0 \left(k_S r\right)j_2 \left(k_S r\right)$ and $D\left(k_S r\right)=3h_2\left(k_s r\right)-2\left(\widehat{\rho}-1\right)h_0\left(k_s r\right)$. Here $h_{0,2}$ and $j_{0,2}$ are spherical Hankel and Bessel  functions, and $k_S$ is the shear-mode wavenumber. The effective wavenumber becomes a sum of that found through the $LB$ model and an extra-term due to the shear-mode conversion:
\be
\frac{K_{eff}^2}{k_C^2}=\left[\frac{K_{C}^2}{k_C^2}\right]_{LB}+\Delta_{CS}\frac{K_{C}^2}{k_C^2},
\ee
with
\be
\Delta_{CS}\frac{K_{C}^2}{k_C^2}=-\frac{27\phi^2}{\left(k_C r\right)^6}\frac{k_C^3\left(ik_S b\right)}{k_S\left(k_C^2-k_S^2\right)}T_1^{SC}T_1^{CS}X
\label{extraterms},
\ee
and
\be
X=\sum_{m=n+2}=k_C b j_m'\left(k_C b\right)h_m\left(k_C b\right)-k_S b j_m\left(k_C b\right)h_m'\left(k_C b\right)
\ee
where $n=\left\{-2,0\right\}$, $\phi$ is the bubble fraction, and $b=2r$ is the radius of the excluded volume.

\begin{figure*}[t]
\centering
\includegraphics[width=16cm]{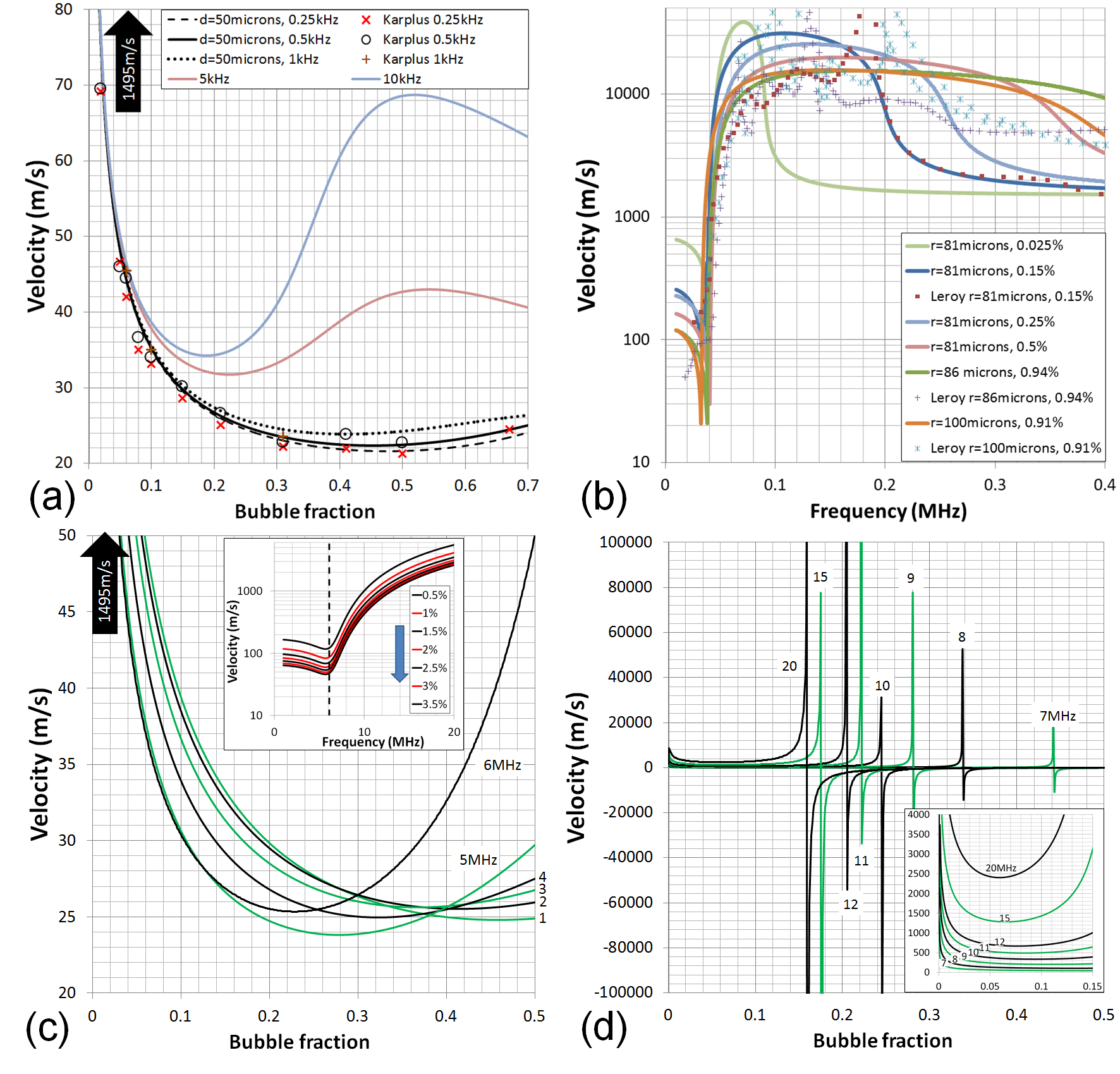}
\caption{The effective phase velocities in bubbly aqueous systems. $(a)$ The effective phase velocity at different bubble fractions with comparison to the experimental data of Karplus \cite{Karplus1958}. $(b)$ Phase velocities for $r=81-100\mu m$ bubbles compared to the experimental results of Leroy et al\cite{Leroy2008}.  In $(c)$ and $(d)$, $r=0.5\mu m$. $(c)$ At frequencies $1-6MHz$ the velocities at different bubble fractions using multiple-scattering theory are shown. The contribution of shear-mode effects is insignificant in a low viscosity continuous phase such as water. The inset shows the velocity-frequency dependence near the Minneart frequency. $(d)$ At higher frequencies, $7-20MHz$, negative effective phase velocities occur. The inset magnifies the view of the velocities found with bubble fraction $\phi=0-0.15$.}
	\label{fgr:AirWater}
\end{figure*}

\begin{figure*}[t]
\centering
  \includegraphics[width=10cm]{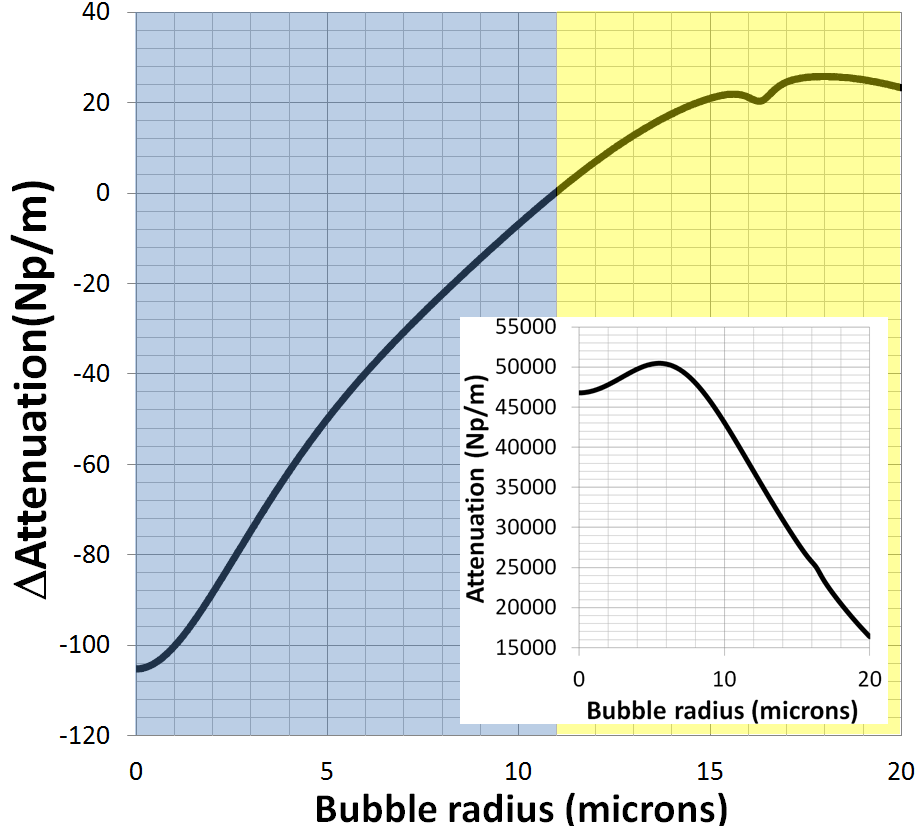}
  \caption{The change in attenuation $\Delta\alpha_{eff}=\alpha_{eff}^{LB}-\alpha_{eff}^{SM}$ at $6.75MHz$ and $20\%$ volume concentration. The shear-mode conversion results in an increase in attenuation below $r=11\mu m$ due to backscatter (blue area). The inset shows the attenuation as a function of bubble radius.  }
	\label{fgr:shearmode2}
\end{figure*}

The propagation of sound through a liquid containing bubbles was discussed in $1947$ by Carstensen and Foldy\cite{Carstensen1947}. The Foldy approach gives reasonable results for low concentrations of bubbles \cite{Leroy2008}. In order to test the multiple-scattering theory that we have developed we make comparison to the work of Karplus\cite{Karplus1958} and of Leroy et al \cite{Leroy2008} - which were based upon bubbles formed in water or weak gels with acoustic properties comparable to water. The latter was an experimental analysis in the frequencies surrounding resonance and the results were modelled using the popular Foldy method. The shear-mode multiple scattering formulation fits well to the results of Karplus (see Fig.\ref{fgr:AirWater} $(a)$) when the bubble radius is $25\mu m$ and the speed of sound in a bubble is $300m/s$. The Karplus analysis was in the acoustic range up to a few $kHz$, well within the audible spectrum and far removed from the resonance frequencies of the bubbly water\cite{resonances}. The objective of the $2008$ work by Leroy and co-workers was to alleviate the scarcity of experimental data\cite{Commander1989} for higher concentrations of bubbles and to test the validity of using the Foldy model. Thus, in Fig.\ref{fgr:AirWater} $(b)$ the results of using the multiple-scattering theory are found to be of matching performance to the Foldy model at concentrations up to $1\%$, as one might expect. Sizes of up to $r=100\mu m$ are analysed with the speed of sound in the dispersed bubbles being $340m/s$ allowing the best fit. The contribution of shear-mode conversion is negligible in the aqueous or aqueous-like low viscosity systems and a $LB$ model alone can suffice. However, these are still relatively low bubble fractions and shear-mode effects become important when the concentration level is high and/or bubble size is small, and the viscosity of the surrounding medium is increased. Keeping to the aqueous continuous phase and decreasing the bubble size to $r=0.5\mu m$ shows some interesting effects, although we find these to be devoid of significant shear-mode effects too. In Fig. \ref{fgr:AirWater} $(c)$ the effective phase velocity in the frequency range $1-6MHz$ is shown, with the inset presenting the velocity/frequency relationship around a Minnaert frequency of $\approx 6.5MHz$. In Fig.\ref{fgr:AirWater} $(d)$ the frequency regime $7-20MHz$ brings about Fano resonance signatures at very high bubble fractions and pronounced negative phase velocities (similar to those found for metafluids \cite{Metafluid2015}). Leroy et al \cite{Leroy2008} found that at lower concentrations the Foldy model performed well around resonance frequencies and so too do the $LB$ multiple scattering models. 
Having confidence that the model works well against experimentally found data an investigation of bubbles formed in oil is now demonstrated. When the frequency of an ultrasonic compressional wave matches the resonance frequency of a bubble a maximum system attenuation is reached. Above the resonance frequencies the attenuation becomes very large, resulting in high effective velocities. Energy transfer is further complicated when considering viscous dissipation in an analysis. When the bubbles are spread out energy loss through shear waves occurs, but in higher concentrations shear-mode reconversion dictates an effective increase or decrease in attenuation and hence speed of sound. This can be most pronounced near a resonant mode. We choose to model bubbles in castor oil, which has a viscosity of $\eta=0.985Pas$. In the supplementary information an investigation of the stability of bubbles in $1)$ an aqueous medium containing $10ppm$ medical grade silver and $2)$ castor oil is shown. In castor oil the bubbles are very stable. 
\begin{figure*}[t]
\centering
  \includegraphics[width=16cm]{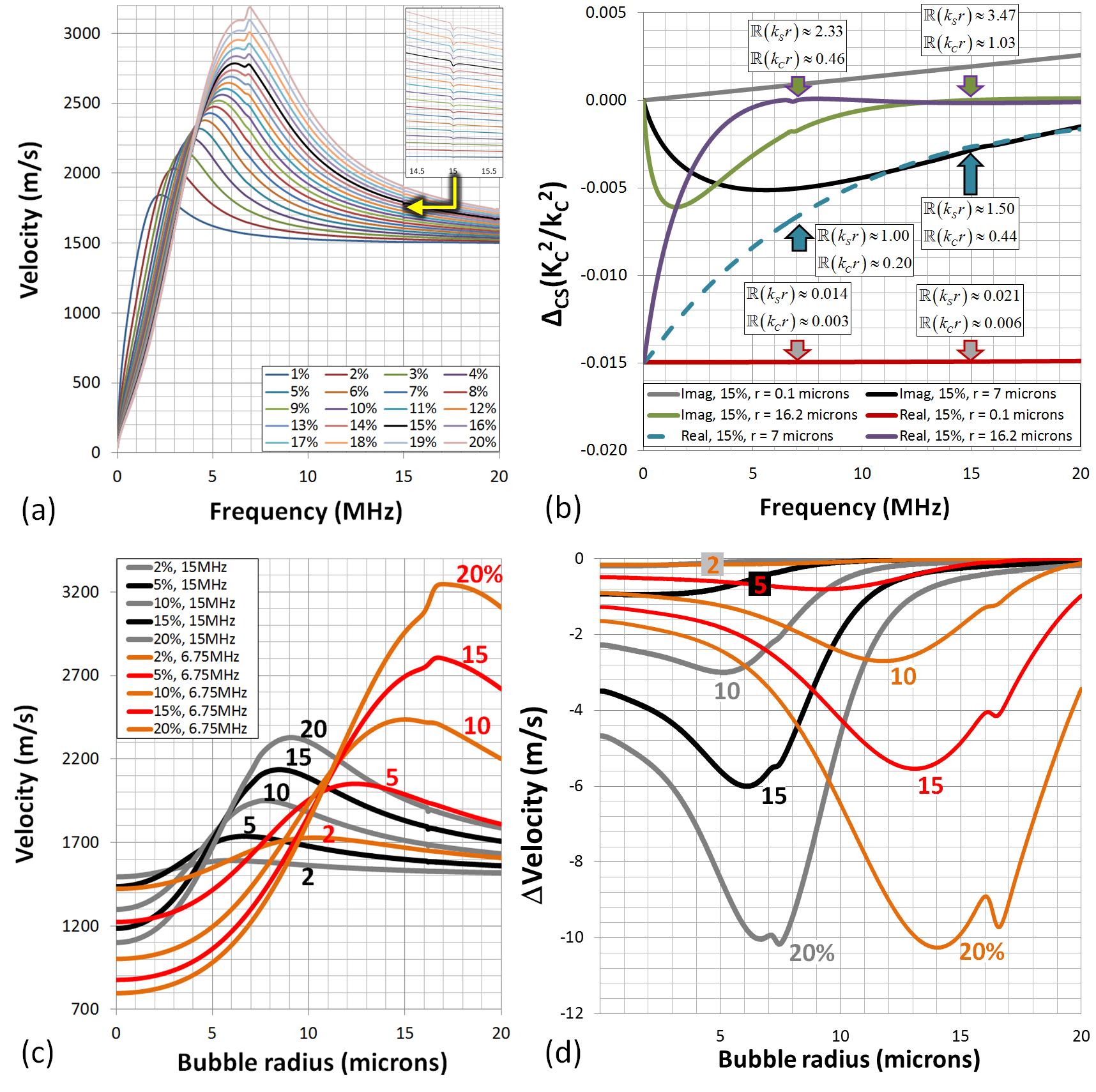}
  \caption{The additional shear-mode conversion changes occurring around the resonance frequencies of microbubbles.  $(a)$ The phase velocities for $r=16.2\mu m$ bubbles at $1-20\%$ volume concentration with a frequency range of $0-20MHz$. $(b)$ The contribution of shear-mode conversion to the effective wavenumber for monodispersed bubbles of radii $1\mu m$, $7 \mu m$, and $16.2\mu m$. $(c)$ The phase velocity as a function of bubble radius at $6.75MHz$ and $15MHz$ in the concentration range $2-20\%$. $(d)$ The change in effective phase velocity due to shear-mode effects compared to a Lloyd-Berry multiple scattering approach.}
	\label{fgr:shearmode}
\end{figure*}
Figure \ref{fgr:shearmode} $(a)$ presents the phase velocities at bubble concentrations of $1-20\%$, for frequencies up to $20MHz$ in a mono-dispersion of $r=16.2\mu m$ bubbles. This particular bubble radius was chosen because in the additional shear-mode real and imaginary parts (see Fig.\ref{fgr:shearmode} $(b)$) there appears two small dips at $6.75MHz$ and $15MHz$, which can be clearly seen to relate to sharp changes in the effective phase velocity. In Fig.\ref{fgr:shearmode} $(b)$ the extra shear-mode terms of Eq.(\ref{extraterms}) are plotted as a function of frequency and some indication of $\Re(k_C r)$ and $\Re(k_S r)$ given for bubble radii of $0.1\mu m$, $7\mu m$, and $16.2\mu m$. One can note that there are no extra resonances found for $r=0.1\mu m$ and $r=7\mu m$ and it can be seen from Fig.\ref{fgr:shearmode} $(c)$ that they are unique to radii $\approx 16.2\mu m$. With the higher viscosity of oil, as compared to water, the contribution of shear-waves in the bubbly liquid can no longer be neglected when the concentration becomes greater than $\approx1\%$. Figures \ref{fgr:shearmode} $(c)$ and $(d)$ illustrate the contributions due to shear-effects in terms of the effective phase velocities (related to the real parts of Eq.(\ref{extraterms}) and shown in Fig.\ref{fgr:shearmode} $(b)$). In $(c)$ the velocities at different concentrations of bubbles are shown at $6.75MHz$ (orange/red) and $15MHz$ (grey/black) for different bubble radii. Taking the difference in velocities found between a standard $LB$ model and the new shear-mode ($SM$) conversion one, $\Delta c_{eff}=c_{eff}^{LB}-c_{eff}^{SM}$, shows that the phase velocity increases due to $SM$ effects as a function of size/concentration and peaks around a resonance frequency. The signature of additional resonances is clear in the double minima of Fig.\ref{fgr:shearmode} $(d)$ and most obvious at $20\%$ concentration. At this bubble concentration the difference in attenuation is plotted in Fig.\ref{fgr:shearmode2} at $6.75MHz$. The attenuation is increased for bubble radii smaller than $11\mu m$ in this case, and reduced when $r>11\mu m$. A reduction in the concentration to, for example, $10\%$ results in a positive change in attenuation with $r>9\mu m$. Thus, lowering the bubble fraction can adjust the attenuating properties of the system with regards to the shear mode contribution.  

\section*{Conclusions}
We have shown that for low bubble fractions in water gel that a multiple-scattering approach in the spirit of Lloyd and Berry \cite{LloydBerry1967} is as accurate as any existing model through comparison to the works of Karplus\cite{Karplus1958} and Leroy et al\cite{Leroy2008}. We introduced the shear-mode conversion model (\cite{ForresterNanoscale2016,Metafluid2015} that is predicted to describe viscoelastic systems containing bubbles. The phenomena associated with shear-mode effects occur for relatively high bubble concentrations (greater than $1\%$) and with highest contributions to the phase velocities and attenuations near resonance. Current technology allows the creation of monodispersed bubbles in a liquid using techniques such as capillary dragging\cite{Leroy2008}, microfluidic flow focusing\cite{Garstecki2004}, and fluidic oscillation\cite{Zimmerman2011}. Analyses of high concentrations of bubbles as a function of bubble size are important in ultrasonics because of the desire to create novel contrast agents for medicine\cite{Zha2013}, nano- and micro-bubbles for sonoluminescence experiments and applications\cite{FlintSuslick1991,Rivas2012}, cleaning and sterilisation, and even metamaterials\cite{Bretagne2011}. Thus, with a technology drive for use of ultrafine bubbles under-way it is inevitable that higher concentrations of bubbles will be produced for novel new industrial methodologies and the work herein demonstrates that shear-mode conversion can possibly occur with a significant contribution to the overall effective wavenumber. 

\section*{Supplementary Information}
An investigation of the stability of microbubbles is undertaken that shows that longer bubble lifetimes occur by forming clusters in a low viscosity colloidal dispersion of silver nanoparticles in water. Clustering of particles has been shown to have a strong influence on the ultrasonic attenuation \cite{Coupled2018,FEMemulsions2018,FEMsolids2019}.  In a more viscous continuous phase, such as castor oil, streams of micro/nano bubbles are then seen to remain stable even in relative isolation. The bubbles that are smaller than $10 \mu m$  are very robust. Thus, shear-mode conversion effects should be observable in bubbly dispersions.
\begin{figure*}[t]
\centering
  \includegraphics[width=15cm]{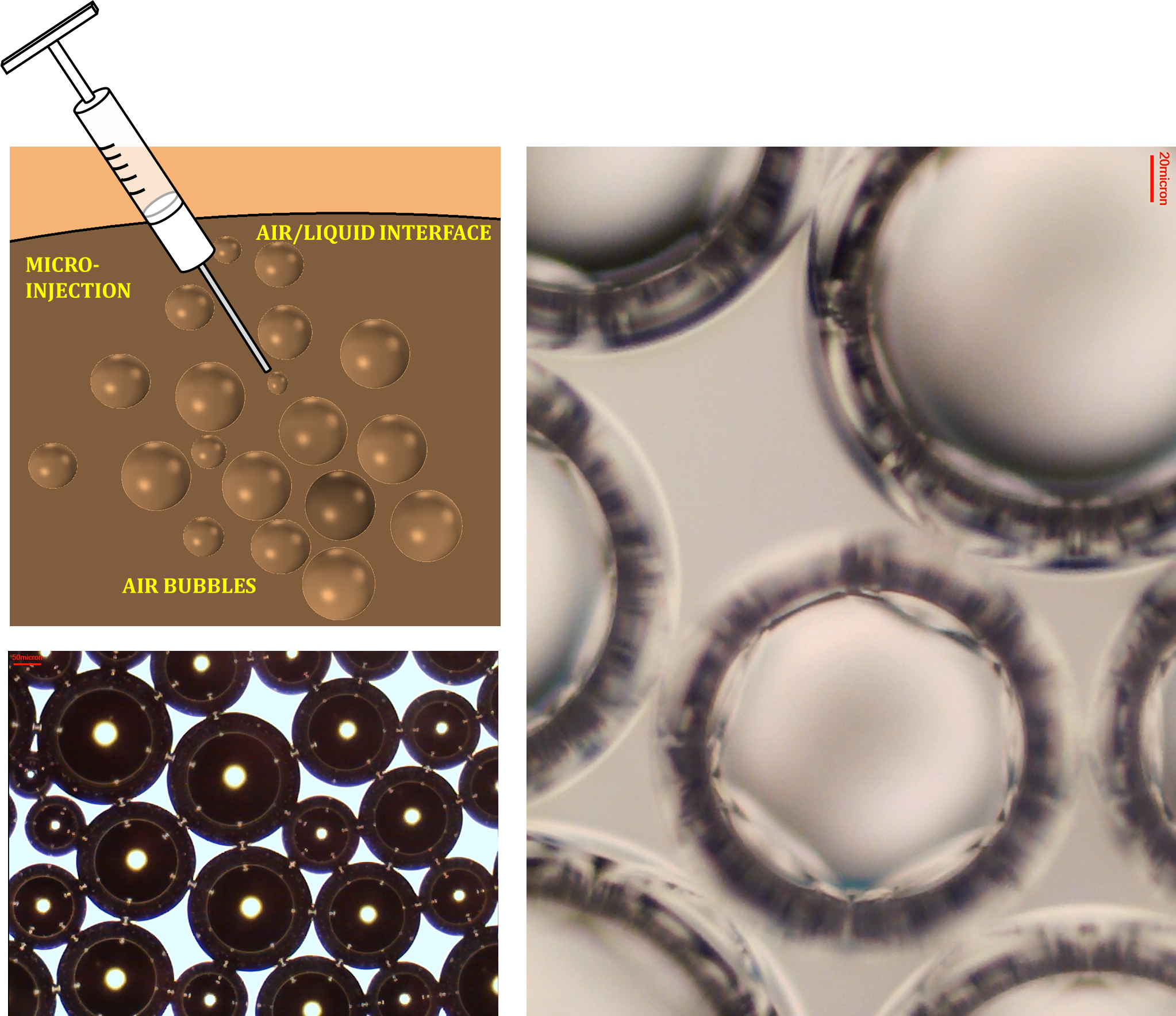}
  \caption{Microbubbles in an aqueous suspension.}
	\label{fgr:S1}
\end{figure*}
Prior to investigating ultrasonic attenuation in bubbly dispersions the stability of different bubble size distributions is addressed. In this work we examine the experimentally observed interaction of bubbles in highly contrasting continuous phases: 1) an aqueous medium containing 10ppm medical grade silver ($0.6-5 nm$), and 2) castor oil derived from the ricinis communis plant. The bubbles are generally small compared to the acoustic wavelengths.  Figure \ref{fgr:S1} shows that upon injection of bubbles into the water/silver suspension using a micro-syringe that the bubbles form clusters. These clusters are strong attractors to other neighbouring bubbles and this draws them together.    
\begin{figure*}[t]
\centering
  \includegraphics[width=15cm]{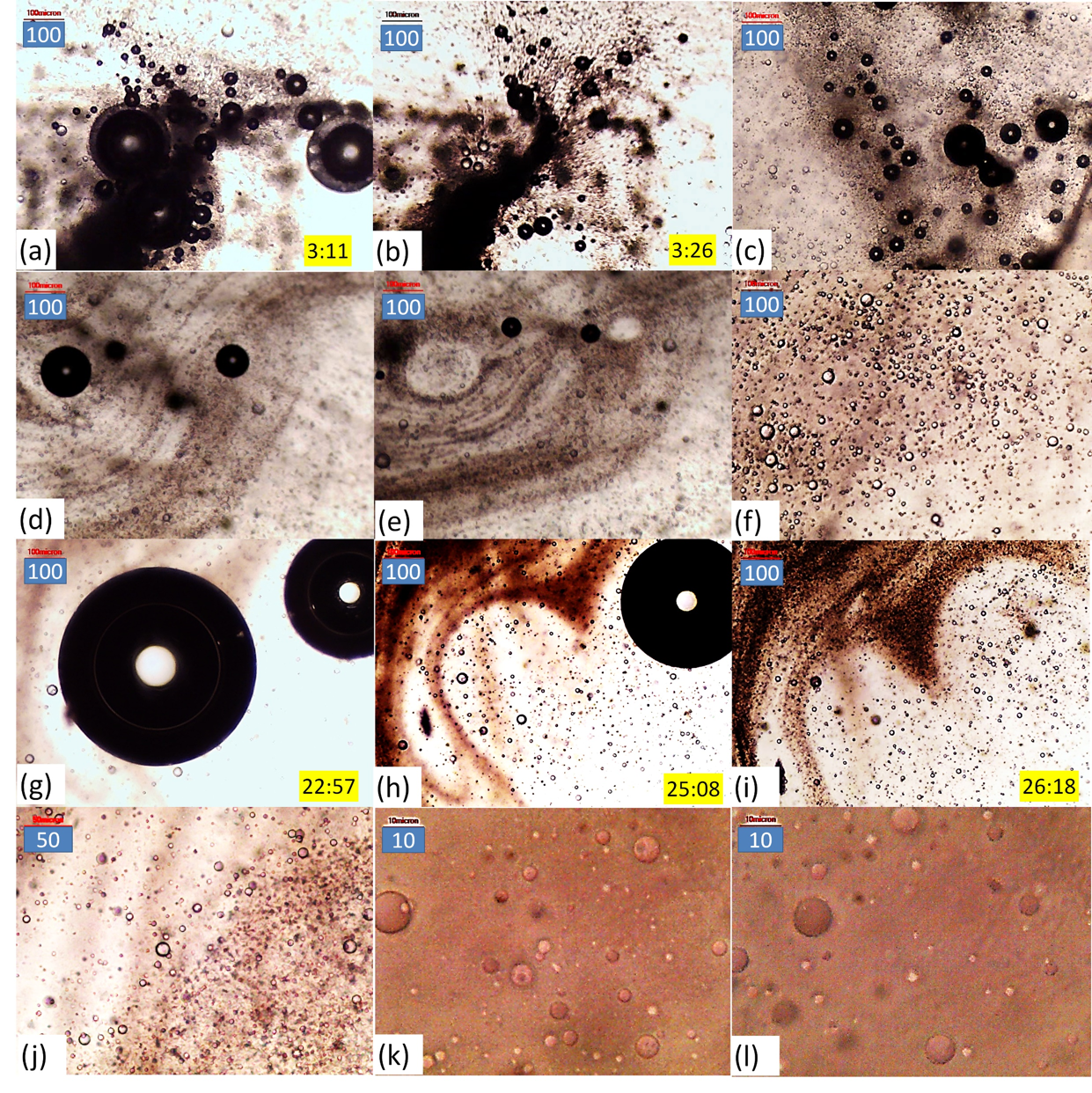}
  \caption{Air bubbles in castor oil. Scale bars in microns are shown in the top-left corner of  each snap-shot.}
	\label{fgr:S2}
\end{figure*}
The finding that bubbles cluster to form stability is also supported by the simulations of Weijs, Seddon, and Lohse \cite{stable2012}.  The stability of bulk nanobubbles and the reason for slow dissolution  was proposed to be a close packing between bubbles ($<10r$ between centres). In Fig. \ref{fgr:S1} the cluster is very tight and smaller bubbles can diffuse into larger ones through Ostwald ripening. 
Figure  \ref{fgr:S2} shows that micro-injection causes streaks of bubbles in castor oil, with some stable and some under continual rising motion. The bubbles were injected at high speed. The larger bubbles eventually explode leaving their mark in the distribution of the surrounding bubbles. In Fig.  \ref{fgr:S2} (a) and (b) one can see that between 3 minutes, 11 seconds and 3 minutes 26 seconds some large $\approx 100 \mu m$ bubbles have collapsed. Throughout the sample the vigorous injection of bubbles left a highly polydispersed system with large bubbles seemingly less stable than the smaller ones. In (c) – (d) one can see that eventually most large diameter bubbles have disappeared and sub $10 \mu m$ ones remain. Further injection shows the same pattern as the concentration increases. For example, between $22:57$ and $26:18$, in (g) – (h), the collapse of the two very large bubbles leaves a quasi-static contour of smaller bubbles around where they once existed. Figure \ref{fgr:S2} (j) –(l) shows the remaining stable formations, with sub-$1 \mu m$  bubbles clearly present in the dispersion. Even several weeks later the smaller bubbles still existed in this viscous continuous phase. Further experiments designed to create monodispersed distributions of bubbles for low intensity ultrasonic testing in viscoelastic media are in development and will be reported elsewhere.

\end{document}